# Scalable Sources of Entangled Photons with Wavelength on Demand

Rinaldo Trotta,[1]* Javier Martín-Sánchez,[1] Johannes S. Wildmann,[1] Giovanni Piredda,[3] Marcus Reindl,[1] Christian Schimpf,[1] Eugenio Zallo,[2] Oliver G. Schmidt,[2] Sandra Stroj,[3] Johannes Edlinger,[3] and Armando Rastelli[1]

[1]Institute of Semiconductor and Solid State Physics, Johannes Kepler University Linz, Altenbergerstr. 69, A-4040 Linz, Austria.

[2]Institute for Integrative Nanosciences, IFW Dresden, Helmholtzstr. 20, D-01069 Dresden, Germany.

[3]Forschungszentrum Mikrotechnik, FH Vorarlberg, Hochschulstr. 1, A-6850 Dornbirn, Austria

**Corresponding Author**

Rinaldo Trotta

Institute of Semiconductor and Solid State Physics,

Johannes Kepler University Linz,

Altenbergerstr. 69, A-4040 Linz, Austria.

Tel.: +43 732 2468 9599

Fax: +43 732 2468 8650

e-mail: rinaldo.trotta@jku.at

The prospect of using the quantum nature of light for secure communication keeps spurring the search and investigation of suitable sources of entangled-photons. Semiconductor quantum dots are arguably the most attractive. They can generate indistinguishable entangled-photons deterministically, and are compatible with current photonic-integration technologies, a set of properties not shared by any other entanglement resource. However, as no two quantum dots are identical, they emit entangled-photons with random energies. This hinders their exploitation in communication protocols requiring entangled-states with well-defined energies. Here, we introduce scalable quantum-dot-based sources of polarization-entangled-photons whose energy can be controlled via dynamic strain-engineering without degrading the degree of entanglement of the source. As a test-bench, we interface quantum dots with clouds of atomic vapours, and we demonstrate slow-entangled-photons from a single quantum emitter. These results pave the way towards the implementation of hybrid quantum networks where entanglement is distributed among distant parties using scalable optoelectronic devices.

The success of semiconductors in nowadays information and communication technology is tightly linked to two aspects of classical semiconductor devices: miniaturization and scalability. While miniaturization is not an issue also for devices operating in the quantum realm, scalability has – hilariously – remained a neglected promise. The reason is that nanostructures used either as storage or sources of quantum bits are not identical, a property arising from the limited capability to fabricate nanostructures with atomic-scale precision[1]. A prominent example is represented by semiconductor quantum dots (QDs)[2]. During the radiative decay of a confined biexciton (label with XX in Fig. 1), these nanostructures can generate entangled photon-pairs[3,4,5,6] with high efficiency[7,8], high degree of entanglement[8,9,10], high indistinguishability[11,12], and - in contrast to parametric down conversion[13] and four-wave mixing[14] sources – they can deliver photons deterministically[11,15]. Compared to other single quantum emitters[16], QDs have the undeniable advantage of being compatible with the mature semiconductor technology[17,18]. For these reasons, it has been recently argued that they have the potential to become the "perfect" sources of entangled photons[19]. In spite of these accomplishments there are two points that are often largely overlooked: First, in the presence of structural asymmetries of the QDs the anisotropic electron-hole exchange interaction induces a fine structure splitting FSS ($s$) between the intermediate exciton X levels[20] that dramatically lowers the degree of entanglement of the source[21], which eventually emits only classically polarization-correlated photons (see Fig. 1a). Even though there exist methods to suppress the FSS[6,22], advanced quantum optics experiments[23] are still carried out using single "hero" QDs that have – for probabilistic reasons[24] – zero FSS. Second, even having at hand a bunch of these special QDs, each of them emits entangled-photons at a different random energy, and any attempt to modify these energies via external perturbations restores the FSS, thus spoiling entanglement[9]. This hurdle hinders the use of dissimilar QDs in quantum optics experiments that require entangled-states with identical energies[25]. This random "pinning" of the emission energy also prevents the possibility of photon storage in quantum memories based on atomic clouds[26], as the existing protocols require precise matching of the photon energy with atomic resonances.

Here, we show that the solution to these problems can be found in the source of the problems itself, i.e., in the solid state nature of such nanostructures and on their strong coupling to mechanical deformations. Specifically, the core idea of our work is to manipulate the strain status of the QD and surrounding semiconductor matrix so as to achieve full control over the anisotropic electron-hole exchange interaction[20], and to modify the energy levels involved in the generation of entangled photons (X, and XX, see Fig. 1a) without opening the FSS (see Fig. 1d). Addressing this task is not trivial. Theory and experiments have demonstrated that the X level degeneracy can be restored via the combined action of two independent external perturbations, such as stress and electric fields[6]. Suppression of the FSS ($s=0$) occurs, however, for a particular combination of the magnitude of the two fields and, as a consequence, for a specific and unpredictable energy of the emitted photons[9]. Although it has been observed that for some special QD tuning of the photon energy at small, albeit non zero, FSS can be achieved using a combination of magnetic and electric fields[27], the demonstration of an energy-tunable source of entangled photons is still lacking. Recently, some of us have theoretically proposed that independent control over the three components of the QD in-plane strain tensor would allow the development of energy-tunable sources of entangled photons[28]. We now present the experimental implementation of this theoretical concept. To benchmark our results we precisely tune a QD ("artificial atom") to emit entangled photons in the spectral region between double absorption resonances of natural atoms (see Fig. 1e).

Our device (see Fig. 1b) is built up merging piezoelectric and semiconductor technologies[29]. The self-assembled In(Ga)As QDs studied here are embedded in the free-standing area of a 300 nm-thick GaAs nanomembrane that is bonded onto a micromachined 300 µm-thick $[Pb(Mg_{1/3}Nb_{2/3})O_3]_{0.72}[PbTiO_3]_{0.28}$ piezoelectric substrate (PMN-PT). The actuator features six trapezoidal areas ("legs") separated by trenches aligned at ~ 60° with respect to each other (for details about the device fabrication, see the supplementary material). The key idea behind this design is that full control of the in-plane strain tensor can be achieved by applying *three independent uniaxial stresses* in the nanomembrane plane. With our actuator, quasi-uniaxial stresses in the membrane can be obtained by applying three independent voltages ($V_1$, $V_2$, $V_3$) at the bottom of opposite legs (labelled as Leg 1,2,3) with respect to the top part of the piezo, which is electrically grounded. Fig. 1c shows a top-view microscope picture of the central part of the device. The regions of the nanomembrane that are suspended can be clearly distinguished from those that are bonded on the piezo-legs.

In order to modify the energy of the entangled photons emitted during the XX-X-0 radiative cascade (see Fig. 1d), it is first necessary to achieve full control over the FSS. To do so, external fields with *two* different degrees of freedom are required to master *two* different QD parameters[28]: the magnitude of the FSS (*s*), and the polarization direction of the exciton emission (θ). The latter parameter gives the in-plane orientation of the exciton states with respect to the crystal axis[30], and it provides information about the QD anisotropy that are fundamental to drive the device during the experiment. A robust approach[6] to achieve zero FSS with two "tuning knobs" can be illustrated by picturing the QD anisotropy as an ellipse with axes given by the two in-plane spin-spin coupling constants[20] (see Fig 2a): the FSS can be suppressed every time one external field is used to align θ along the direction of application of the second field, which is then capable to compensate completely for the asymmetries in the in-plane QD confining potential, i.e., it is capable to bring the FSS through zero (see Fig. 2a). Following this picture, the first step we performed in the experiment is a polarization-resolved measurement aimed at measuring $s$ and θ when no voltages are applied to the piezo-legs. These two quantities are encoded in polar plots of Fig. 2b, where the length and orientation of the "petals" give the value of the $s$ and θ, respectively (see Methods). The measurement at zero applied voltages (see the black data-points in Fig. 2b) reveals $s$=20 ± 0.3 µeV and θ=109°±0.4° with respect to the [110] direction of the GaAs nanomembrane, which was aligned perpendicularly to Leg 2 during device fabrication. Since θ differs from the direction of application of the stress exerted by Leg 2 (labelled $\phi_2$ in the figure) by 19°, it is not possible to suppress the FSS using only this Leg. In fact, Fig. 2e shows clearly that sweeping the voltages across Leg2 leads to an anticrossing between the two X states[31], with a lower bound for the FSS of ~ 10 µeV. The X degeneracy can instead be restored when we first use Leg 1 to align θ along Leg 2 (see red data points in Fig. 2b), and then we employ Leg 2 to cancel the FSS. Fig 2e shows that when θ~$\phi_2$ the FSS can be tuned well below the threshold of 1 µeV usually required to observe entangled photon emission (see the black-data points on the left-hand side of Fig. 2a). These results represent the first experimental evidence that the FSS can be fully controlled with two independent stresses, in line with earlier theoretical calculations[22]. We observe this behaviour in all eight QDs we selected randomly in two different devices, thus proving the general relevance of our findings. It is important to note that, for each QD, there is only one combination of $V_1$ and $V_2$ that allow the condition *s*=0 to be reached, similarly to the case of strain and electric field[9]. To implement an energy-tunable source of entangled photons, the X energy ($E_x$) has to be modified without re-opening the FSS. Intuitively, this can be achieved by adding an isotropic biaxial stress to the QD, which does not affect its in-plane anisotropy (see Fig. 2a). To accomplish this task we make use of the third piezo leg: By applying a voltage $V_3$ across it we modify the strain configuration and thus the initial values of s and θ (see the blue data-points in Fig. 2c). We then follow the two-step procedure described above to erase the FSS. (1) We use Leg 1 to rotate θ until the exciton aligns along Leg 2 or the perpendicular direction, i.e., until we reach the condition θ~$\phi_2$ (see the supplementary material). The specific direction is determined by the handedness of θ under the stress exerted by Leg 1, being clockwise (anticlockwise) when is larger (smaller) than 120˚. (2) Finally, we suppress the FSS by sweeping again the voltage on Leg 2 (see Fig. 2e). Having modified the QD strain status with Leg 3, the condition *s* = 0 is now obtained for a different combination of $V_1$ and $V_2$ and, as a consequence, for a different X energy emission (see the black-data points on the right-hand side of Fig. 2a). It is worth mentioning that the role of the legs can be exchanged without affecting the final results, as there exists only one combination of the three in-plane components of the stress tensor which leads to *s*=0 at each specific X energy[28]. To change $E_X$, it is then sufficient to repeat the three step procedure described for a different combination of $V_1$, $V_2$ and $V_3$. This is made easier considering that the voltages $V_1$ and $V_3$ to be applied to maintain the condition $\theta \sim \phi_2$ scale linearly (see Fig. 2d). Finally, Fig. 2f shows that at $s=0$ $E_X$ changes linearly with $V_2$ and, most importantly, that it can be tuned across a spectral range of ~ 7 meV. The experimental results shown so far are fully in line with a theoretical model based on **k·p** theory that describes the behavior of the X states under the influence of in-plane strains with variable magnitude and anisotropy[28]. This model further confirms that our six-legged device is capable of delivering three independent stresses to single QDs (see the supplementary material), the key ingredient to reach the results shown in this work.

Having demonstrated for the first time that it is possible to tune the energy of the X at $s=0$, we now demonstrate that our device can be used as an *energy-tunable source* of entangled photons. For this experiment, we choose a *different QD* and we repeat the three step procedure detailed above by carefully tuning the FSS down to zero with a resolution of ~ 0.2 μeV (see Fig. 3a). For two different X energies (highlighted with $E_{X1}$ and $E_{X2}$ in Fig. 3a) we performed polarization-resolved cross-correlation measurements between the X and XX photons (see methods). In the case $s=0$, the two-photon state can be expressed by the maximally entangled Bell state $\psi = (|R_{XX} L_X\rangle + |L_{XX} R_X\rangle)/\sqrt{2}$, which can be equivalently rewritten as $\psi = (|H_{XX} H_X\rangle + |V_{XX} V_X\rangle)/\sqrt{2}$ or $\psi = (|D_{XX} D_X\rangle + |A_{XX} A_X\rangle)/\sqrt{2}$, where *H* (*V*), *D* (*A*) and *R*(*L*) indicate horizontally (vertically)-polarized, diagonally (anti-diagonally)-polarized, and left (right)-circularly polarized photons, respectively. Therefore, when performing polarization-resolved cross-correlation measurements, one expects a strong bunching peak for co-linear, co-diagonal, and cross-circular polarization and antibunching for the opposite polarization settings. This is exactly the behaviour we observe in our measurements (see Fig. 3b-c), which represent a clear signature of entanglement. In order to estimate the degree of entanglement, we calculate the fidelity $f$ to the Bell state $\psi$ (see methods) and we found – by integrating over 0.5 ns around the central peak – $f_1 = 0.80\pm0.05$ and $f_2 = 0.78\pm0.06$ for $E_{X1}$ and $E_{X2}$, respectively (see Fig. 3d-e), similarly to what we found in different QDs by full reconstruction of the two photon density matrix[9]. These values are larger than the classical limit (0.5) for a source emitting polarization-correlated photons, and prove for the first time that QDs can produce entangled states of light at different energies. The fact that the two values of the fidelity integrated over the time delay are identical within the errors further implies that the degree of entanglement of the source does not depend on the photon energy, an important requirement for applications[32]. Despite the fidelity we measure is among the highest ever reported so far for QDs[8,9], it is not yet perfect due to depolarization of X states[33] and recapture processes[7]. However, we strongly believe that the level of entanglement can be even improved further using faster photon detectors and resonant two-photon excitation schemes[11,15].

In order to prove that we are able to control the energy of the entangled photons with the precision required for advanced quantum optics experiments, we show that it is possible to interface entangled photons emitted by a QD with clouds of natural atoms operated as a slow-light medium. Pioneering works on the field[34,35] have demonstrated that single photons emitted by GaAs QDs can be slowed down using warm rubidium vapors, which can be also exploited as absolute energy-reference to interconnect the different nodes of a quantum network. The concept has been recently adopted to interface single photons emitted by single molecules with vapors of alkali atoms[36]. However, it has never been applied to slow-down entangled photon-pairs from single quantum emitters. The reason is that the energy of one photon of the entangled-pair must match the spectral window of double absorption resonances in warm atomic vapours, and this possibility was out of reach before our work.

The emission spectrum of our QDs is centred at around 900 nm, close to the D1 lines of Cesium (Cs). Therefore, we insert a temperature-stabilized quartz cell containing Cs vapour[37,38] in the optical path of the exciton (see methods). Fig. 4a shows a sketch of the Cs D1 lines of relevance for this experiment, i.e., the transitions involving the $6^2P_{1/2}$ level and the hyperfine-split $6^2S_{1/2}$ doublet[39]. In order to observe slow-entangled-photons, one has to tune the energy of the X (or XX) transition exactly in between the hyperfine lines[34] while *s* is kept at 0 (see black and purple arrows in Fig. 4a). Experimentally, this condition is identified by recording the intensity of the X emission while scanning its energy across Cs (see Fig. 4b). This allows us to observe two absorption resonances resulting from a convolution between the inhomogenously-broadened X transition and the Doppler-broadened Cs doublet (see the supplementary material). When the X energy is tuned exactly in the middle of the hyperfine lines ($E_{X3}$, see Fig. 4b and Fig. 3a), the QD photons probe a medium with a strong variation of the refractive index with frequency and are slowed-down[34]. Taking into account the temperature of the Cs cell used in the experiment ($T_{Cs}$= 140 ˚C), we estimate a differential delay of up ~2 ns on a 7.5 cm long path. To observe this delay, it is sufficient to perform X-XX cross-correlation measurements with and without the Cs cell. Fig. 4c shows the result of such measurements when both X and XX photons are collected with the same diagonal polarization. The bunching peak clearly shifts by more than 1.8 ns when the Cs cell is inserted into the optical path. However, the peak appears less pronounced and broader compared to the measurements performed without Cs cell. It is therefore interesting to investigate whether the degree of entanglement of photon-pairs is affected by the presence of the atomic cloud. Fig. 4d shows the fidelity calculated after performing XX-X cross-correlation measurements with the same polarization settings used in Fig. 3b-c. The measurements without Cs cell show a peak-fidelity of $f_3^{off}=0.8\pm0.02$, which nicely matches the values obtained for the other $E_{X1}$ and $E_{X2}$ (see Fig. 3) and further confirms that the degree of entanglement of our source does not change with the X energy. The measurements with the Cs cell show instead a peak fidelity of $f_3^{on}=0.65\pm0.08$. Despite this value is above the classical limit of 0.5, it may suggest a degradation of the degree of entanglement. However, careful inspection of the data shows that this is not the case and that the peak fidelity is not the most accurate parameter to estimate entanglement. The solid lines in Fig. 4d show lorentzian fits to the experimental data. In the presence of the Cs cloud there is a considerable temporal broadening (a factor almost 2), while the integrated area (A) remains practically unchanged ($A_{off}/A_{on}=1.06$). These results clearly imply that the degree of entanglement of the photons is not affected by the presence of the Cs cell. On the other hand, the Cs cell affects the temporal distribution of the transmitted photons because of the strongly dispersive behaviour combined with the spectral broadening of the X emission line, which we estimate to be ~ 35 µeV (see Fig. 4b, the supplementary material and Ref. 38).

In conclusion, we have demonstrated for the first time that it is possible to modify the energy of the polarization-entangled photons emitted by QDs without affecting the degree of entanglement of the quantum emitters. These results have been achieved by developing a novel class of electrically-controlled semiconductor-piezoelectric devices that allow the optical and electronic properties of single QDs to be arbitrarily re-shaped via anisotropic strain engineering. The tunability of the photon-energy has opened up the unprecedented possibility to interface entangled photons from QDs with clouds of natural atoms and, in turn, to demonstrate for the first time slow-entangled photons from a single quantum emitter. In light of our results, it is possible to envisage a new era for QDs in the field of large distance quantum communication. In fact, dissimilar QDs can now be used for entanglement teleportation (or swapping), as the entangled-photons can be color-matched for quantum interference at the beam splitter[25]. Furthermore, the hybrid artificial-atomic interface we have built up can be extended up to the limit where entangled-states are stored and retrieved at the single photon level[26]. This could allow the whole concept of a quantum repeater[40] to be implemented. Addressing all these applications most likely requires the use of our energy-tunable entangled-photon source in combination with photonic structures capable to boost the flux of QD photons[7,8, 41,42] and with resonant excitation schemes for the deterministic generation of entangled photons with high degree of indinshushability[11,15]. However, the implementation of the perfect source of entangled photons is worth the efforts, as the realization of a solid-state based quantum network for the distribution of quantum entanglement among distant parties[43] will be revolutionary.

**METHODS**

**Sample growth and device fabrication.** In(Ga)As QDs were grown by molecular beam epitaxy. Following oxide desorption and buffer growth, a 100 nm-thick $Al_{0.75}Ga_{0.25}As$ sacrificial layer was deposited before a 300 nm-thick GaAs layer containing the QDs. The QDs were grown at 500 °C and capped by an indium flush technique. Using standard epoxy-based photo-resist, the sample (coated with a 100 nm-thick Cr/Au layer) is integrated via a flip-chip process onto a Cr/Au-coated micro-machined PMN-PT actuator. The GaAs substrate, the buffer layer, and the AlGaAs sacrificial layer were removed using wet chemical etching (see the supplementary material), thus leaving the GaAs nanomembrane tightly bound on the piezoelectric actuator. The PMN-PT actuator was processed in the six-leg design shown in Fig. 1 with a femtosecond laser having central wavelength of 520 nm, pulse duration of 350 fs, repetition rate of 25 KHz and a ~5 μm spot size. Six metal contacts were fabricated at the bottom of the piezo-legs so that independent voltages can be applied with respect to the top contact, which is set to ground. The same voltage is applied to opposite legs to limit displacements of the central structure. The voltage applied on each pair of legs leads to a well-controlled deformation of the GaAs nanomembrane suspended in the central part (see the supplementary material). Differently from previous works, we pole the piezoelectric actuator so that a voltage V<0 (V>0) applied to each pairs of aligned legs result in an out-of-plane electric field that leads to an in-plane contraction (expansion) of the piezo-leg. Therefore, an in-plane tensile (compressive) strain is transferred to the central part of the nanomembrane (see supplementary material). In the tensile regime (V<0), the X emission line of the QD of Fig. 2 shifts approximately linearly with the applied voltages, with a slope that slightly depends on the leg used. In particular, we measure slopes of 10 µeV/V, 7 µeV/V, 3 µeV/V for Leg 1, Leg 2, and Leg 3, respectively. This difference most probably arises from the fact that the QD under investigation is not sitting in the very central part of the device (see the supplementary material). However, we cannot exclude different bonding conditions between the piezo-legs and the GaAs nanomembrane.

**Micro-photoluminescence and photon-correlation spectroscopy.** Conventional micro-photoluminescence (PL) spectroscopy is used for the optical characterization of the devices. The measurements are performed at low temperatures (typically 4-10 K) in a helium flow cryostat. The QDs are excited non-resonantly at 840 nm with a femtosecond Ti:Sapphire laser and focused by a microscope objective with 0.42 numerical aperture. The measurements in Fig. 3 and Fig. 4 appear to be performed with a continuous wave laser due to long lifetime of the X transition. The same objective is used for the collection of the PL signal, which is spectrally analysed by a spectrometer and detected by a nitrogen-cooled silicon charge-coupled device. Polarization-resolved micro-photoluminescence experiments are performed combining a rotating half-wave plate and a fixed linear polarizer placed after the microscope objective. The transmission axis of the polarizer is set parallel to the [110] direction of the GaAs crystal (within 2°) and perpendicular to the entrance slit of the spectrometer, which defines the laboratory reference for vertical polarization. The FSS and the polarization angle of the X emission are evaluated using the same procedure reported in Ref. 6, which ensures sub-microelectronvolt resolution.

For photon-correlation measurements, the signal is split into two parts after the microscope objective using a non-polarizing 50/50 beam splitter, spectrally filtered with two independent spectrometers tuned to the XX and X energies, and finally sent to two Hanbury Brown and Twiss setups (HBTs). Each HBT consists of a polarizing 50/50 beam splitter placed in front of two avalanche photodiodes (APDs), whose output is connected to a 4-channel correlation electronics for reconstructing the second-order cross-correlation function between the XX and X photons, $g^{(2)}(\tau)$. The temporal resolution of the set-up is about 500 ps, mainly limited by the time jitter of the APDs. Properly oriented half-wave plates and quarter-wave plates are placed right after the non-polarizing beamsplitter to select the desired polarization. With a single measurement, this experimental setup allows $g^{(2)}_{XX,X}(\tau)$ to be reconstructed in 4 different polarization settings, so as to minimize the effect of possible sample drifts. The count rate at each APD is typically $10^3$–$10^4$ count per seconds and the integration time used for each measurement is about 1 hour.

For cross-correlation measurements in the presence of the atomic clouds, we use a 7.5 cm long quartz cell filled with Cs vapours and wounded up with heating foils for precise temperature tuning. The temperature chosen for the experiment is 140±0.1 °C. The Cs cell is inserted in the X optical path, after the non-polarizing beam splitter and the retarder wave-plates, right before the entrance slit of the spectrometer. When the X is tuned in between the hyperfine lines of Cs, the count rate at the APDs is reduced by a factor ~ 2 due to the optical absorption in the Cs vapour.

**Entanglement analysis.** Raw data were used to evaluate the second order correlation functions, without any background light subtraction. As mentioned in the previous section, the experimental setup allows the second order XX-X cross-correlation function $g_{AB}^{(2)}(\tau)$ in 4 different polarization settings (AB) to be evaluated with a single measurement, i.e., $g_{AB}^{(2)}(\tau)$, $g_{BA}^{(2)}(\tau)$, $g_{AA}^{(2)}(\tau)$, $g_{BB}^{(2)}(\tau)$ are measured simultaneously. Having 4 measurements for each polarization basis has two main advantages, it compensates for measurement drift and it allows us to calculate the degree of correlation without any assumption on the polarization state of the photons. In particular, we do not need to assume, e.g., that $g_{AA}^{(2)}(\tau) = g_{BB}^{(2)}(\tau)$, and we can simply average the observed results, i.e., $g_{AA}^{(2)} = (g_{AA}^{(2)} + g_{BB}^{(2)})/2$. The degree of correlation is then calculated using the following formula $C_{AB} = (g_{AA}^{(2)} - g_{AB}^{(2)})/(g_{AA}^{(2)} + g_{AB}^{(2)})$. Finally, the fidelity is calculated via[17]

$$f = (1 + |C_{HV}| + |C_{DA}| + |C_{RL}|)/4.$$

**Figure 1. A six-legged semiconductor-piezoelectric device for quantum optics**. **(a)**. Sketch of the radiative decay of a confined biexciton (XX) to the ground state (0) in a generic as-grown QD. In the presence of a fine structure splitting FSS (*s*) between the two bright exciton states X, the emitted photons are only classically correlated. H (V) indicates horizontally (vertically) polarized photons. **(b)** Sketch of the six-legged device used to engineer the strain status of a GaAs nanomembrane (grey region) containing single QDs. The nanomembrane is integrated onto a micro-machined PMN-PT piezoelectric actuator. The whole device is placed onto a patterned chip-carrier which allows independent voltages to be applied to the bottom of the piezo-legs. **(c)**. Microscope picture of the central part of the final device. The same voltage (V) is applied to pairs of opposite legs to limit displacement of the central structure. **(d)**. Same as in (a) for a QD embedded in the device show in (c), where anisotropic in-plane strains are first used to restore the X degeneracy for the generation of entangled-photons (left panel), and then to modify the X and XX energies without affecting the FSS (right panel). The yellow lines indicate entanglement. $\sigma^+$ ($\sigma^-$) indicates right (left) circularly-polarized photons. **(e)**. Sketch of a QD whose X photon – polarization-entangled with the XX photon – is tuned to the middle of the hyperfine levels of Cs and it is slowed down. The Cs vapour is contained in a quartz cell.

**Figure 2**. **Tuning the exciton energy at zero fine structure splitting**. **(a)**. Sketches of the in-plane QD anisotropy under the effect of external perturbations, see the main text. The major and minor axes represent qualitatively the two in-plane spin-spin coupling

constants, see Ref. 20. **(b)**. Dependence of ΔE as a function of the angle the linear polarization analyser forms with the [110] crystal axis, where ΔE is half of the difference between the XX and X energy minus its minimum value (see methods). The length and orientation of the petals give the value of the *s* and θ, respectively. The black data-points correspond to zero applied voltages to the piezo-legs, while the red data-points show the configuration when Leg 1 is used to achieve θ ~ $\phi_2$. The solid lines are sinusoidal fits to the experimental data. The [110] crystal direction of the GaAs nanomembrane is also indicated. **(c)**. Same as in (b) when Leg 3 is used to change the QD strain status (see the blue data-points) while Leg 1 is again used to achieve θ ~ $\phi_2$ (see the red data-points). The curve for zero applied voltages is also reported for reference (see the black data points). **(d)**. Linear dependence of the voltage $V_1$ applied to Leg 1 vs. the voltage $V_3$ applied to Leg 3 when θ~ $\phi_2$. **(e)**. Behaviour of the FSS as a function of the energy of the X for a single QD embedded in the device shown in Fig. 1c. The X energy is varied by sweeping the voltage ($V_2$) on Leg 2. The voltages on Leg 1 ($V_1$) and Leg 3 ($V_3$) are tuned so that the polarization direction of the exciton emission (θ) is tuned with respect to the direction of application of Leg 2 ($\phi_2$), as described in the text. The dashed line shows the threshold of 1 µeV usually required to observe highly entangled photons during the XX-X-0 cascade. **(f)**. Behaviour of the X energy as a function of the voltage $V_2$ applied to Leg 2 at *s*=0.

**Figure 3. Entangled photon pairs at different photon energies**. **(a)**. Same as in Fig. 2e for a different QDs whose FSS is fine-tuned to the spectral resolution of the experimental set-up (~ 0.2 µeV). The circles indicates the X energies ($E_{X1,2,3}$) where cross-correlation

measurements have been performed. In particular, $E_{X3}$ correspond to the position in between the hyperfine lines of Cs. **(b)**. XX-X cross-correlation measurements for diagonal (top panel), circular (central panel) and linear (bottom panel) polarization basis when the X energy is set to $E_{X1}$. **(c)**. Same as in (b) for $E_{X2}$. **(d)**. Fidelity to the maximally entangled Bell state ψ as a function of the time delay when the X energy is set to $E_{X1}$. The dashed line indicates the threshold for entanglement, i.e., the classical limit. **(e)**. Same as in (d) for $E_{X2}$.

**Figure 4. Entangled photons from artificial atoms interfaced with natural atoms**. **(a)**. Sketch of the energy levels of a Cs atom (left panel) and of a QD fine-tuned for entangled photon generation (right panel). The energy axis (not to scale) has been shifted so as to highlight that the energy of the X transition (purple arrow) matches exactly the middle of the two D1 lines of Cs (black arrows). The energy of the XX transition is indicated by the orange arrow. **(b)**. Intensity of the X transition when its energy is swept across the D1 lines of Cs. The solid line represents a fit to the theoretical model (see the supplementary material) used to extract the linewidth of the X transition (indicated with $FWHM_{QD}$). The dashed arrow indicates the energy $E_{X3}$ where cross-correlation measurements have been performed with and without the Cs cell. **(c)**. Cross-correlation measurements for X and XX photons with the same diagonal polarization with (red line) and without (blue line) the Cs cell in the X optical path. The temporal delay introduced by the Cs cell (τ) is also indicated. **(d)**. Fidelity to the maximally entangled Bell state ψ between X and XX photons with (red line) and without (blue line) the Cs cell in the X optical path. The solid lines are lorentzian fits to the experimental data. The ratio between the full-width-half maxima (*w*) of the curves and the integrated area (A) are also indicated.

ACKNOWLEDGMENT

We thank G. Katsaros, G. Bauer, A. Predojecvic, C. Ortix, and F. Schäffler for fruitful discussion and T. Lettner and D. Huber for help. The work was supported financially by the European Union Seventh Framework Programme 209 (FP7/2007-2013) under Grant Agreement No. 601126 210 (HANAS), and the AWS Austria Wirtschaftsservice, PRIZE Programme, under Grant No. P1308457.

AUTHOR CONTRIBUTIONS

R. T. conceived and designed the experiment. R. T. performed measurements and data analysis with help from J. M. S., J. S. W., M. R. and A. R.. J. M. S. processed the device with help from C. S., R. T., and A. R.. G. P., S. S., and J. E. processed the piezoelectric-actuators, designed by J. M. S. and A. R.. E. Z. and O. G. S. performed MBE growth. R. T. wrote the manuscript with help from all the authors. R. T. and A. R. coordinated the project.

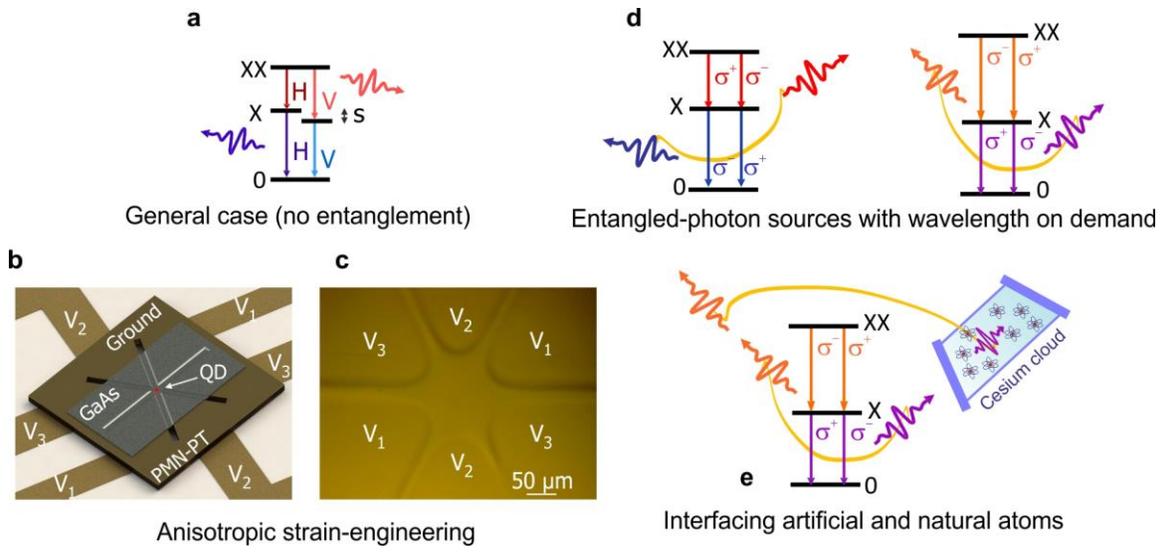

Figure 1 of 4
by Trotta, R. *et al*.

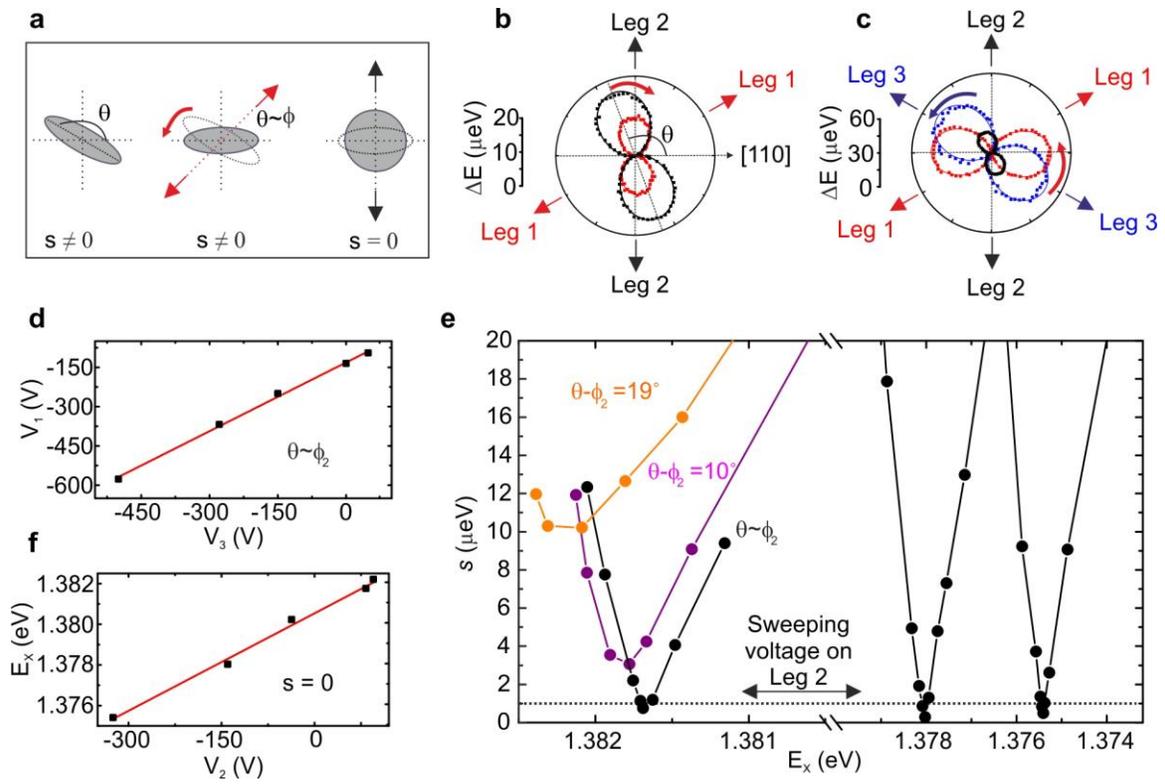

Figure 2 of 4
by Trotta, R. *et al*.

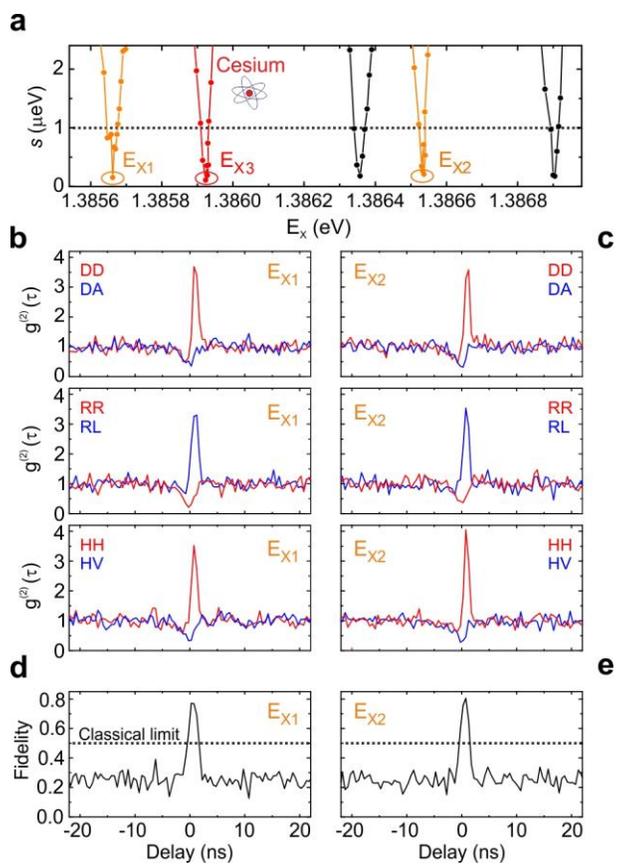

Figure 3 of 4
by Trotta, R. *et al*.

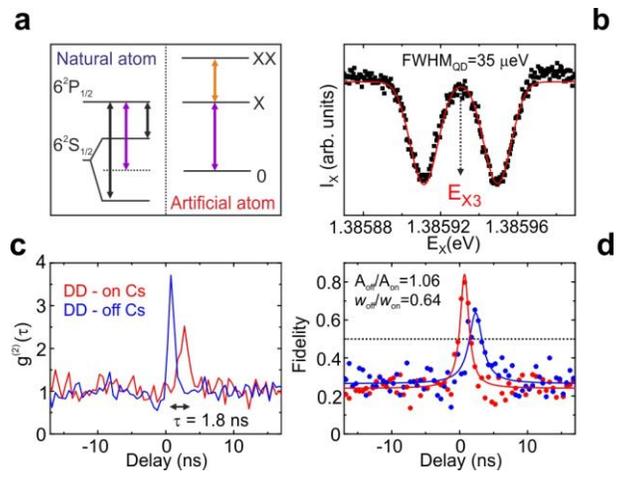

Figure 4 of 4
by Trotta, R. *et al*.

[30] Gong, M., Zhang, W., Guo, G.-C. & He, L. Exciton Polarization, Fine-Structure Splitting, and the Asymmetry of Quantum Dots under Uniaxial Stress. Phys. Rev. Lett. **106**, 227401 (2011)

[31] Bennett, A. J., Pooley, M. A., Stevenson, R. M., Ward, M. B., Patel, R. B., Boyer de la Giroday, A., Sköld, N., Farrer, I., Nicoll, C. A., Ritchie, D. A., Shields, A. J. Electric-field-induced coherent coupling of the exciton states in a single quantum dot. Nat. Phys. **6**, 947 (2010).

[32] Ekert, A. K. Quantum cryptography based on Bell's theorem. Phys. Rev. Lett. **67**, 661 (1991).

[33] Stevenson, R. M.; Salter, C. L.; Boyer de la Giroday, A.; Farrer, I. A.; Nicoll, C. A.; Ritchie, D. A.; Shields, A. J. Coherent entangled light generated by quantum dots in the presence of nuclear magnetic fields. ArXiv:1103.2969.

[34] Akopian, N., Wang, L., Rastelli, A., Schmidt, O. G., & Zwiller, V. Hybrid semiconductor-atomic interface: slowing down single photons from a quantum dot. Nature Phot. **5**, 230 (2011).

[35] Akopian, N., Trotta, R., Zallo, E., Kumar, S., Atkinson, P., Rastelli, A., Schmidt O. G., & Zwiller, V. An Artificial Atom Locked to Natural Atoms. arXiv:1302.2005 (2013)

[36] Siyushev, P., Stein, G., Wrachtrup, J., & Gerhardt, I. Molecular photons interfaced with alkali atoms. Nature **509**, 66–70 (2014).

[37] Ulrich, S. M.. Weiler, S., Oster, M., Jetter, M., Urvoy, A., Löw, R. & Michler, P. Spectroscopy of the D1 transition of cesium by dressed-state resonance fluorescence from a single (In,Ga)As/GaAs quantum dot. Phys. Rev. B **90**, 125310 (2014).